# MgB$_2$ superconductor thin films on Si and Al$_2$O$_3$ substrates

A. Plecenik[1], L. Satrapinsky[1], P. Kúš[2], Š. Gaži[1], Š. Beňačka[1], I. Vávra[1] and I. Kostič[3]

[1] *Institute of Electrical Engineering, Slovak Academy of Sciences, Dubravska cesta 9, 84239 Bratislava, Slovak Republic*
[2] *Department of Solid State Physics FMFI, Comenius University, Mlynska dolina, 84215Bratislava, Slovak Republic*
[3] *Institute of Informatics, Slovak Academy of Sciences, Dubravska cesta 9, 84237 Bratislava, Slovak Republic*

**Abstract**

Thin films of MgB$_2$ superconductor were prepared by three different procedures on sapphire and silicon substrates. Boron thin films, ex-situ annealed in magnesium vapour, resulted in textured polycrystalline films with crystal dimensions below about 1μm, onset critical temperature T$_{con}$ near 39 K and width of phase transition ΔT ≤ 1 K. Both, ex-situ and in-situ annealed co-deposited boron and magnesium thin films on sapphire and silicon substrates give smooth nanocrystalline films. DC properties of nanocrystalline MgB$_2$ films co-deposited on silicon substrate reached T$_{con}$ = 33 K and zero resistance T$_{co}$ = 27 K, the highest values received until now on Si substrates. In addition, microwave analyses prove the existence of superconducting parts of the film below 39 K. This result confirms the possibility to synthesise nanocrystalline superconducting MgB$_2$ thin films on silicon substrate with critical temperature near 39 K, prepared by vacuum co-deposition of boron and magnesium films.

## Introduction

The onset critical temperature T$_{con}$ ≈ 39 K of MgB$_2$ superconductor has generated great interest because it reaches the highest critical temperature in the simple two component non-oxide compound [1]. Thin films of MgB$_2$ prepared on technological dielectric or semiconductor substrates represent large promise for electronic applications, including cryoelectronic structures with weak links, due to larger coherence length, much smaller anisotropy and normal state resistivity, and higher compound stability than at present occurs in high temperature multicomponent oxide superconductors. However, high volatility of magnesium and small decomposition temperature of MgB$_2$ creates difficulties at preparation of smooth, homogeneous and epitaxial thin films.

Promising results were received by pulse laser deposition (PLD) from target composed of a mixture of Mg/MgB$_2$ powders [2,3], stoichiometric MgB$_2$ target [4], Mg+B precursor, or by e-beam evaporation of boron [5], and subsequent MgB$_2$ synthesis by ex-situ (~ 900 $^{o}$C) or in-situ (~600 $^{o}$C) annealing procedure. Generally, it was shown that until now thin film synthesis by ex-situ annealing processes indeed gives T$_{con}$ ≈ 39 K [6] but these processes are not suitable for obtaining smooth and (eventually) epitaxial films. On the other hand, in-situ annealing allows to receive smooth nanocrystalline films, however, the onset critical temperature is relatively low (below ~27 K [2,6]). The reason is probably in oxygen contamination, formation of higher phases or not ideal stoichiometry and annealing process. Nevertheless, it seems that in-situ process is necessary for successful preparation of smooth thin films by the above mentioned deposition techniques. Recently, in-situ synthesised $MgB_2$ thin films on sapphire substrate, deposited by PLD, enabled to receive the highest $T_{co} \approx 34$ K [7].

In the paper we present properties of $MgB_2$ thin films deposited on sapphire and silicon (100) single crystal substrates by ex-situ annealing of boron film in magnesium vapour, and by co-deposition of Mg-B film and ex-situ or in-situ $MgB_2$ synthesis. The films of thickness about 200 nm were deposited on room temperature substrates from two resistive heaters. After certain optimisation of the annealing process $T_{con} = 33$ K and $T_{co} = 27$ K was obtained for both, sapphire and Si substrates. The $MgB_2$ films prepared by co-deposition are very smooth and convenient for patterning and preparation of microstrip structures and weak links.

**Experimental**

Three different procedures were applied for the preparation of $MgB_2$ thin films on sapphire and silicon substrates. For the morphological and microstructural studies of prepared films Hitachi S-800 field emission scanning electron microscope (SEM) and JEOL 1200EX transmission electron microscope (TEM) were used. The electrical properties of the films were studied using standard dc four probes method and microwave characterization was realized by a resonator method in 10 GHz frequency band.

In the first technological procedure boron thin film (~150 nm thick) was thermally evaporated on randomly oriented sapphire substrates succeeded by ex-situ annealing in magnesium vapor. The boron thin film was then closed in a Nb tube together with Mg chips in the ratio of Mg:B more than 2:1. The Nb tube was closed to an annular furnace and kept on Ar atmosphere at the pressure ~ 3 kPa. The sample temperature was then increased from room temperature to 800 °C in 60 minutes, 30 minutes was kept on the same temperature and after this the sample was quenched to the room temperature in five minutes. The time vs. annealing temperature illustrates diagram in the Fig. 1a.

The morphology of $MgB_2$ films prepared by this procedure illustrates Fig. 2a. These thin films are composed from superconducting randomly oriented single crystals, with dimensions below about 1 μm, protruding from a continuous bottom film. The beginning of transition to the superconducting state appears at temperature $T_{con}$ close to 39 K, and zero resistance critical temperature $T_{c0} \approx 37$ K (Fig.2b).

The second procedure of $MgB_2$ thin films preparation was realized by co-deposition,

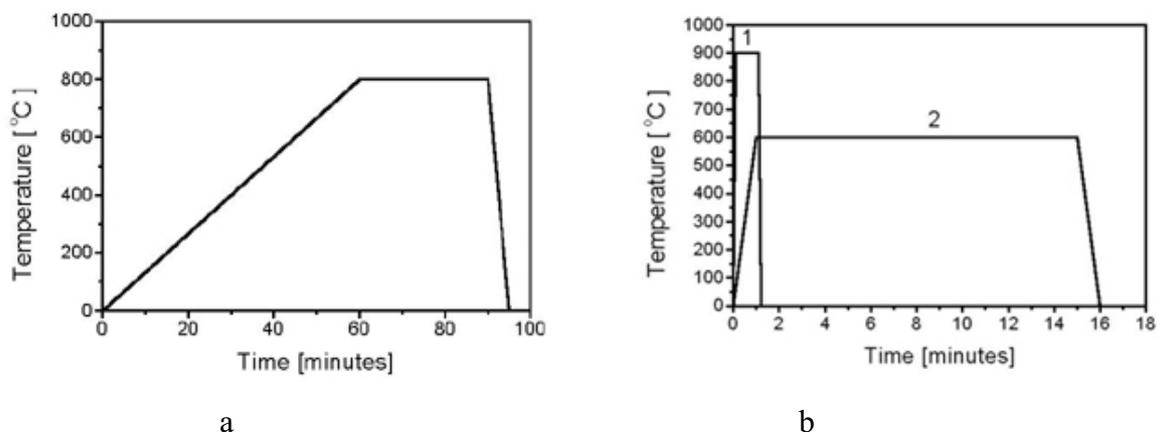

a            b

Fig.1 Time vs. annealing temperature for $MgB_2$ film synthesis: ex-situ annealing in magnesium vapour (a), vacuum in-situ annealing of co-deposited Mg-B films (b, curve 1), and ex-situ annealing in Ar atmosphere (b, curve 2).

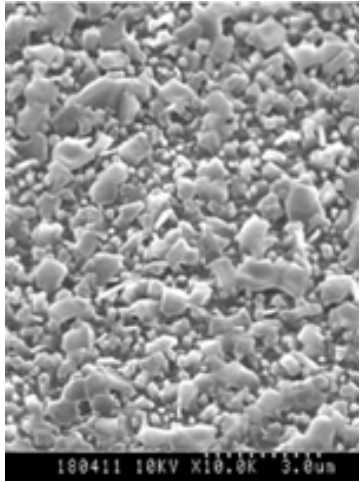
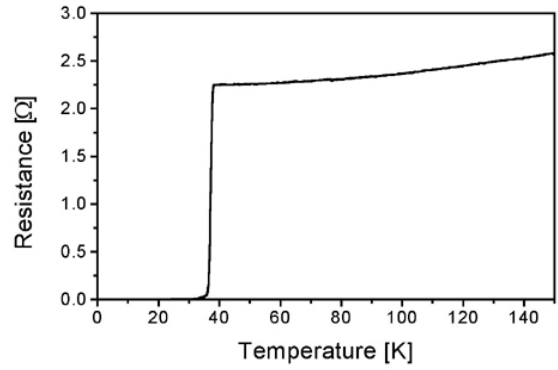

a    b

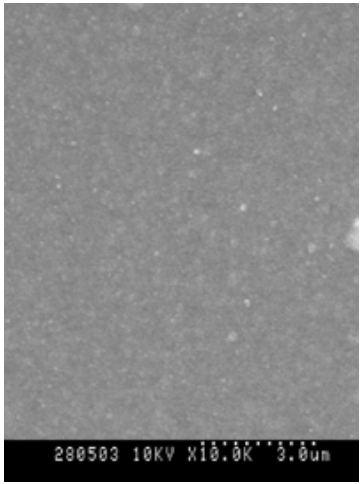
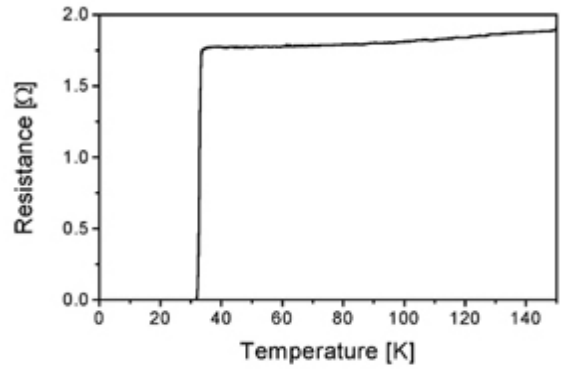

c    d

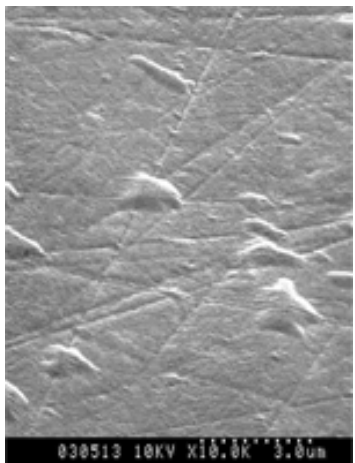
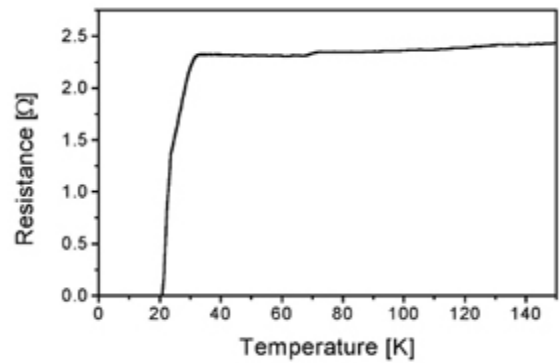

e    f

Fig.2 The morphology and R(T) dependence of the MgB$_2$ thin films prepared on randomly oriented sapphire substrates by: ex-situ annealing of boron thin film in magnesium vapour (a, b), ex-situ annealing in Ar atmosphere of co-deposited Mg-B film (c, d), and in-situ annealing in vacuum of co-deposited Mg-B thin film (e, f).

i.e. by simultaneous evaporation of boron and magnesium from two separate resistive heaters at a vacuum $8.10^{-4}$ Pa. The magnesium and boron films were evaporated, as in the previous procedure, on unheated randomly oriented sapphire substrates. A total thickness of Mg-B thin film was varied from 100 to 200 nm. The ratio of Mg:B composition was held above 2:1, but no influence of higher Mg:B ratio on the critical temperature of $MgB_2$ thin films was observed. Probable reason of this result is evaporation of excessive magnesium from the film surface. This Mg-B film was ex-situ annealed in Ar using temperature-time diagram shown in Fig.1a. The surface is smooth without a visible defects or cracks (Fig.2 c). The R(T) dependence of this $MgB_2$ thin film on sapphire substrate exhibited $T_{con}$ = 33.3 K and $T_{c0}$ = 32 K.

The third procedure of the $MgB_2$ thin films synthesis on sapphire substrates included in-situ annealing of co-deposited Mg-B film, in accordance with the diagram shown in Fig.1b, curve 1. After co-deposition of Mg-B film the temperature of the heater was increased in vacuum to the value 900 $^{o}$C in 20 seconds. The samples were kept on this temperature 30 seconds and then quenched to the room temperature in 20 seconds. Samples prepared by this procedure give $T_{con} \approx 26$ K and $T_{c0} \approx 16$ K. The surface of these thin films was very smooth (Fig.2e, visible lines are scratches on the substrate surface).

The above elaborated co-deposition process, resulting in the smooth thin films, was applied to prepare $MgB_2$ thin films on silicon (100) substrates. The Mg-B films were in-situ annealed in vacuum a short time at 900 $^{o}$C (diagram of Fig.1b, curve 1). The synthesized $MgB_2$ thin films are very smooth with $T_{con} \cong 27$ K and $T_{co} \cong 17$ K, the films contain cracks appearing after thermal cycling (Fig.3a). Sometimes the cracks are visible also in the R(T) dependence (Fig. 3b) as a sharp resistance increase.

To remove this lack the co-deposited Mg-B films were ex-situ annealed 10 minutes, in Ar gas at pressure 1 atm., and several temperatures between 400 $^{o}$C and 630 $^{o}$C. The best result, from the point of view $T_{con}$ and $T_{c0}$, was received at 600 $^{o}$C. The next step of the film optimization was to find an optimal time at 600 $^{o}$C annealing (Fig 4). Two minutes annealing resulted in non superconducting film with metallic R(T) dependence. With increasing of annealing time the $T_{con}$ and $T_{co}$ increased and the highest $T_{con}$ and the smallest width of the phase transition $\Delta T_c$ (between 10 % and 90 % resistance transition) was obtained at 15 minutes annealing ($T_c$ vs. time in Fig.4 and adequate morphology and R(T) dependence are shown in Fig. 3c,d).

An increase of the normal state resistance (above $T_{con}$) was observed if the time of annealing increased (Fig.4, a,b). The samples annealed below 2 minutes at 600$^{o}$C exhibit metallic R(T) dependence and the value of R(120K) was about 1.2 $\Omega$. The R(120K) increased with larger annealing time which improves the superconducting properties (Fig.4b). This can be connected with a partial loss of magnesium from Mg-B thin film, creation of MgO or $Mg_2Si$ at the interface with the Si substrate.

The cross-section of the $MgB_2$ thin film prepared by co-deposition on Si substrate (region I in Fig.5a) and ex-situ annealed is shown on the Fig.5a. The $MgB_2$ thin film (region III) is above bright thin transition layer (region II). The corrugated surface is visible as a region IV. The microstructure of the film was investigated by tranmission electron microscop (TEM) using samples prepared by Ar ion milling. The plane view of microstructure was examined in several levels: at the surface, in the middle of film thickness and at the $MgB_2$/Si interface, using sequential milling from the surface side.

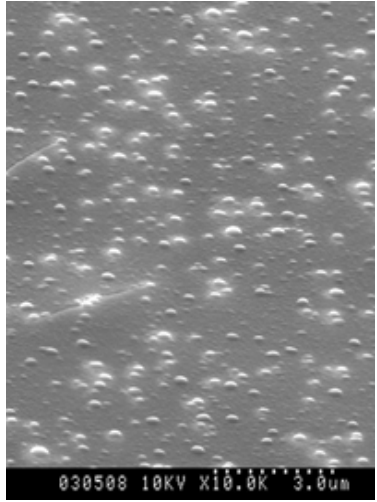
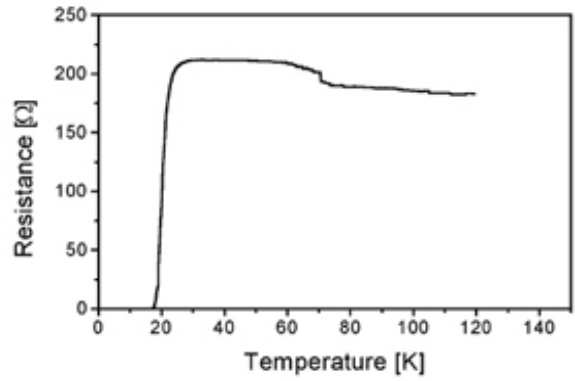

a  b

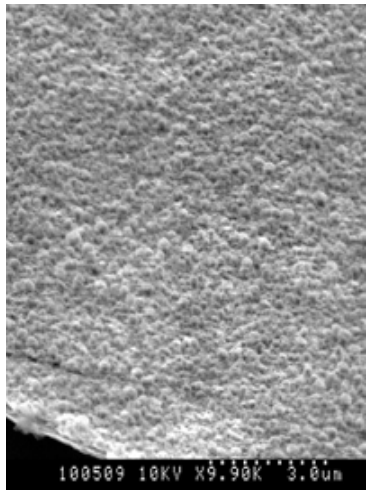
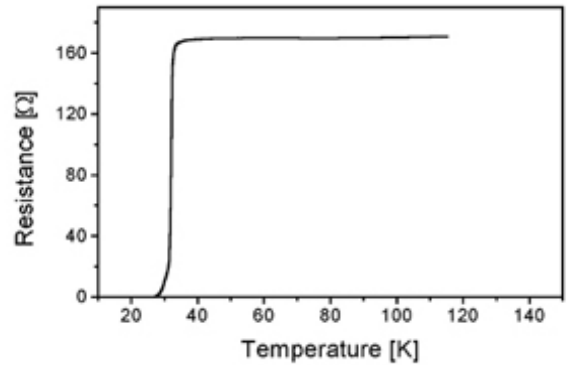

c  d

Fig.3 The morphology and R(T) dependence of the MgB$_2$ superconducting thin films prepared on Si substrates and annealed in vacuum (a,b) or in Ar atmosphere (c,d).

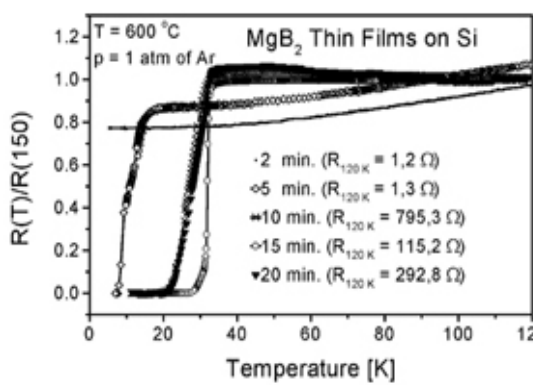
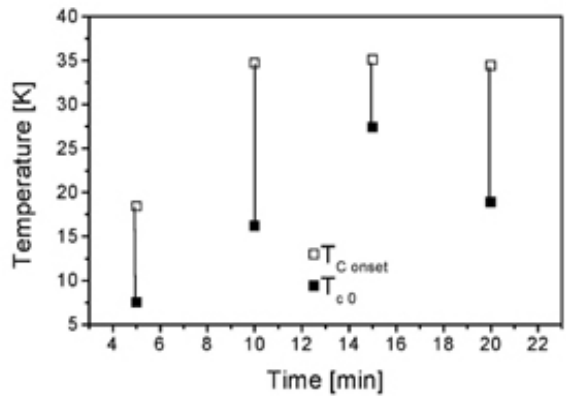

a  b

Fig.4 a) R(T) dependence of MgB$_2$ films, synthesized from co-deposited Mg-B thin films on Si substrate, ex-situ annealed in Ar atmosphere at indicated time. b) The evolution of the onset critical temperature $T_{con}$ and zero resistance critical temperature $T_{co}$ with annealing time.

The electron diffraction (Fig.5b) revealed that the film consists of crystalline and amorphous phases of $MgB_2$. The most of crystalline phase is located at the $Si/MgB_2$ interface, and at the film surface the amorphous phase dominates. The mean grain size is below 15 nm. The TEM investigation of $MgB_2$ thin film at the interface with Si shows a presence of polycrystalline $Mg_2Si$.

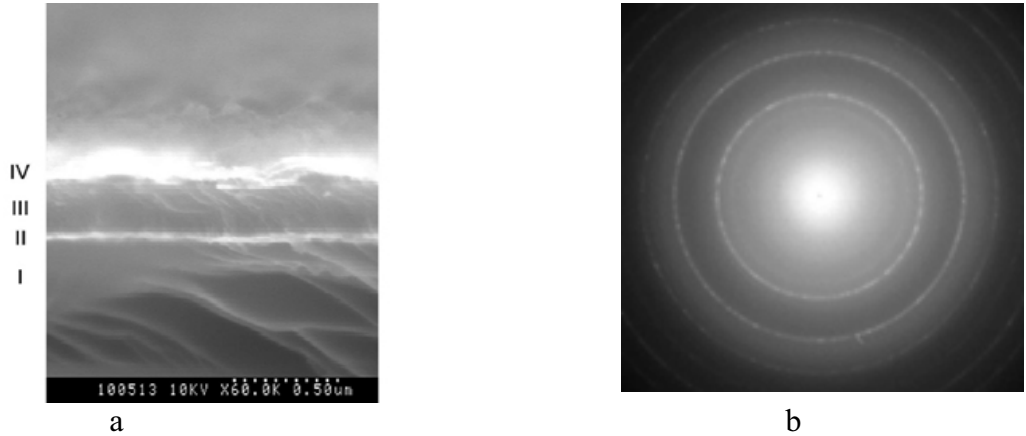

Fig.5 Cross-section (SEM) a) and electron diffraction (TEM) of $MgB_2$ thin film b).

For microwave measurements the two types of the samples were selected – $MgB_2$ prepared on sapphire, shown in Fig.2a, and prepared on Si substrate, shown in Fig.3c. In contrast to dc measurement of R(T) the temperature dependence of microwave losses in the sample can give more complex information about the superconducting properties of the films. We use $TE_{011}$ circular copper resonator filled by sapphire rod with resonant frequency 9.5 GHz and quality factor $Q_o \cong 2000$ without a sample [8]. The temperature dependence of quality factor Q(T) of the resonator filled by the sample characterizes microwave losses in $MgB_2$. Fig.6a shows normalized $Q(T)/Q_o$ at $B_{dc} = 0$ (curve a) and at $B_{dc} = 1$ mT (curve b) measured on the samples prepared on the sapphire substrates (see Fig.2a). The shown change in Q(T) at such small $B_{dc}$ confirms granular character of the sample morphology and a presence of weak link medium, with inter-granular weak links appearing just below the $T_{con}$. This confirms magnetic modulation microwave absorption (MMMA) mode [9] as an additional microwave dissipation δP within the $B_{dc} \cong 0.2$ mT (Fig.6a, lower inset, measured at temperature 37 K). This is in fair agreement with dc R(T) measurements. At lower temperatures (below about 35 K) inter-granular critical currents increase and losses from moving vortices, in dc magnetic field within 4mT perpendicular to the film surface, contribute to microwave losses (upper inset of Fig.6a, T = 24 K). Different properties occur in smooth nanocrystalline $MgB_2$ thin films, deposited on silicon substrate, of the morphology shown on Fig.3c. In spite of $T_{con} \cong 33$ K from dc measurements, microwave quality Q(T) increases below 40 K (Fig.6b) and inter-granular losses (Fig.6b, inset, T = 12 K) appear only at temperature lower than 27 K, the $T_{c0}$ measured by dc method. This result indicate the existence of unconnected superconducting regions at temperature of bulk material and such giving a chance to prepare very smooth thin films on silicon substrates with critical temperature near 39 K.

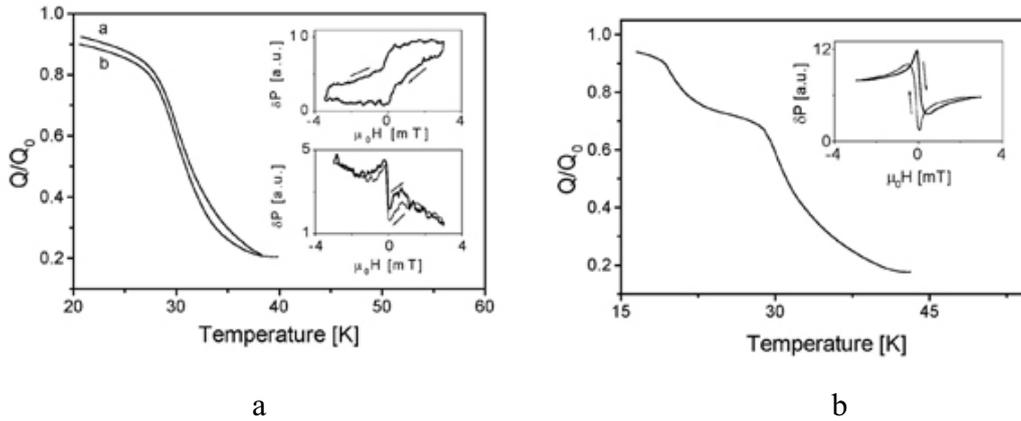

a          b

Fig.6 The temperature dependence of microwave quality factor Q(T) and additional MMMA losses δP(H) from intergranular weak links and moving vortices measured on ex-situ syntetized $MgB_2$, (a) polycrystalline (Fig.2a) and (b) nanocrystalline (Fig.3c), thin films.

In conclusion, $MgB_2$ superconductor thin films prepared by ex-situ annealing in magnesium vapour on sapphire substrate result in polycrystalline films with crystal dimensions in a submicrometer range and critical temperature near the bulk of $MgB_2$. Both, in-situ and ex-situ annealed co-deposited Mg-B films enable to receive smooth nano-crystalline $MgB_2$ films, from transport measurements confirming the $T_{con}$ = 33 K, but containing unconnected superconducting parts with $T_{con} \approx 39$ K measured by microwave MMMA method. This result is promising for preparation of smooth $MgB_2$ thin films on sapphire or silicon substrates, with critical temperature close to the bulk material, by further accomplishing the deposition and annealing processes of co-deposited Mg-B thin films.

**Acknowledgements**

The authors acknowledge the help of D. Machajdik at interpretation of X-ray results and the technical assistance of O.Sekanina. This work has been supported by the Slovak Grant Agency for Science, under the contracts VEGA 2/7199/20, 2/7185/20, 2/7196, and 1/7072/20.